\documentclass[twocolumn,english,aps,prb,superscriptaddress,floatfix]{revtex4}
\usepackage[T1]{fontenc}
\usepackage[latin9]{inputenc}
\usepackage{graphicx}

\makeatletter

\makeatletter

\makeatletter

\makeatletter

\makeatletter

\usepackage{graphics}

\@ifundefined{definecolor}
 {\@ifundefined{definecolor}
 {\@ifundefined{definecolor}
 {\@ifundefined{definecolor}
 {\@ifundefined{definecolor}
 {\usepackage{color}}{}
}{}
}{}
}{}
}{}

\usepackage{bm}

\makeatother

\usepackage{babel}

\makeatother

\usepackage{babel}

\makeatother

\usepackage{babel}

\makeatother

\usepackage{babel}

\makeatother

\usepackage{babel}

\begin{document}

\title{Enhancement of $\mathrm{T_{c}}$ by disorder in underdoped iron pnictide
superconductors}

\author{R. M. Fernandes}

\affiliation{Department of Physics, Columbia University, New York, New York 10027,
USA }

\affiliation{Theoretical Division, Los Alamos National Laboratory, Los Alamos,
NM, 87545, USA}

\author{M. G. Vavilov}

\affiliation{Department of Physics, University of Wisconsin, Madison, Wisconsin
53706, USA}

\author{A. V. Chubukov}

\affiliation{Department of Physics, University of Wisconsin, Madison, Wisconsin
53706, USA}

\date{\today }


\begin{abstract}
We analyze how disorder affects the transition temperature $T_{c}$
of the $s^{+-}$superconducting state in the iron pnictides. The conventional
wisdom is that $T_{c}$ should rapidly decrease with increasing inter-band
non-magnetic impurity scattering, but we show that this behavior holds
only in the overdoped region of the phase diagram. In the underdoped
regime, where superconductivity emerges from a pre-existing magnetic
state, disorder gives rise to two competing effects: breaking of the
Cooper pairs, which tends to reduce $T_{c}$, and suppression of the
itinerant magnetic order, which tends to bring $T_{c}$ up. We show
that for a wide range of parameters the second effect wins, i.e. in
the coexistence state $T_{c}$ can increase with disorder. Our results
provide an explanation for several recent experimental findings and
lend additional support to $s^{+-}$-pairing in the iron pnictides. 
\end{abstract}
\maketitle
\textit{Introduction.}~~The symmetry of the superconducting state
of the iron-based superconductors (FeSCs) is still a subject of intense
debate~\cite{reviews}. Photoemission experiments on moderately-doped
FeSCs show quite convincingly~\cite{ARPES} that the pairing state
is $s-$wave, i.e. fully gapped. However, since the FeSCs are multi-band
systems, the $s-$wave superconducting state (SC) can have either
$s^{++}$ symmetry, if the gaps on different Fermi-surface pockets
have the same sign, or $s^{+-}$ symmetry, if the gaps on different
pockets have opposite signs~\cite{maiti}. The $s^{+-}$ state emerges
due to a repulsive inter-band interaction enhanced by spin-fluctuations
- this is the key element in the theories of a magnetic pairing mechanism
in the FeSCs~\cite{magnetic}. On the other hand, the $s^{++}$ state
emerges if the inter-band interaction is attractive and is enhanced
by orbital fluctuations.~\cite{orbital}

A seemingly straightforward way to distinguish between $s^{+-}$ and
$s^{++}$ pairing symmetries is their responses to impurity scattering.
Both SC states are nearly unaffected by intra-band scattering. However,
while inter-band scattering is harmless to the $s^{++}$ state, it
is pair-breaking to the $s^{+-}$ state, leading to a suppression
of $T_{c}$.\cite{dolgov2009} To verify this experimentally, one
has to choose a dopant that acts predominantly as a non-magnetic impurity
scatterer. In the FeSCs, this is not a trivial task, as many transition-metal
dopants significantly change the carrier concentration.\cite{dhaka11}

One direction explored by many groups was to substitute Zn for Fe.
Early data on LaFeAs(O$_{1-x}$F$_{x}$) showed that $T_{c}$ weakly
depends on the Zn concentration~\cite{Zn_early}, and were interpreted
as an evidence in favor of an $s^{++}$ state. Subsequent studies~\cite{Zn_later,Muromachi10},
however, found that the effect of Zn substitution depends on the doping
level $x$: while in the overdoped regime $T_{c}$ displays a sharp
decrease, in agreement with what is expected for an $s^{+-}$ state,
at optimal doping $T_{c}$ remains virtually the same. More surprisingly,
in the underdoped regime of LaFeAs(O$_{1-x}$F$_{x}$), $T_{c}$ \emph{increases}
with Zn concentration \cite{Zn_later}, an observation that is puzzling
not only for an $s^{+-}$ SC state, but even for a conventional $s^{++}$
state. A similar increase of $T_{c}$ with disorder was found in the
underdoped material Ba(Fe$_{1-x}$Co$_{x}$)$_{2}$As$_{2}$ with
substitution of Cu for Fe \cite{paul}. That the dopant Cu atoms act
as impurity-scatterers follows from both band structure calculations~\cite{dolgov2009}
and neutron scattering experiments. \cite{goldman_Cu} Intriguingly,
measurements in the same Ba(Fe$_{1-x}$Co$_{x}$)$_{2}$As$_{2}$
materials, but with Zn replacing Fe, showed that $T_{c}$ decreases
with increasing Zn concentration even in the underdoped region \cite{Muromachi12},
although $T_{c}$ decreases faster in the overdoped region.

In this paper, we describe the effect of disorder on the superconducting
transition temperature $T_{c}$ of the $s^{+-}$ state in underdoped
samples, when superconductivity develops in the presence of SDW order.
We argue that the conventional wisdom that $T_{c}$ decreases with
increasing impurity concentration does not work in the underdoped
region. Indeed, even in a clean system, the reason why $T_{c}$ goes
down deep in the underdoped region is because SDW order competes with
superconductivity. As doping increases, the SDW order becomes weaker,
and $T_{c}$ increases.\cite{vvc,fs} When disorder is added at a
fixed doping concentration, it influences SDW and SC orders differently:
while both intra-band and inter-band impurity scattering weaken SDW,\cite{kulikov1984}
only inter-band scattering is pair-breaking for $s^{+-}$ superconductivity.\cite{dolgov2009,comm}
As a result, disorder without inter-band scattering component does
not directly affect SC pairing, but weakens the SDW order, leading
to an \textit{increase} in $T_{c}$. The situation is more complicated
when both intra-band and inter-band scattering components are present.
In this situation, impurity scattering affects $T_{c}$ both directly,
via pair-breaking, and indirectly, via the suppression of SDW. Therefore,
the two effects push $T_{c}$ in opposite directions, and whether
$T_{c}$ increases or decreases with increasing impurity concentration
depends on the interplay between the system parameters.

\begin{figure}
\begin{centering}
\includegraphics[width=0.95\linewidth]{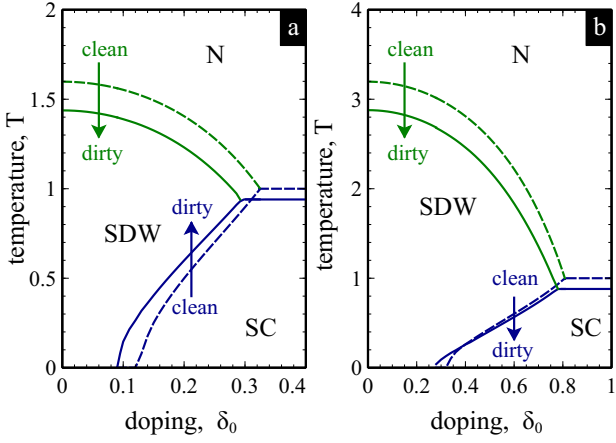} 
\par\end{centering}

\caption{(Color online) Temperatures of normal (N)-to-SDW, SDW-to-SC and N-to-SC
transitions as functions of doping for the clean and dirty cases (dashed
and solid lines, respectively) for two sets of system parameters.
In the underdoped region, where SC emerges from a pre-existing SDW
phase, $T_{c}$ for $s^{+-}$ pairing \textit{increases} with the
concentration of non-magnetic impurities for one set of parameters
(panel a) and weakly \textit{decreases} for the other set (panel b).
These two behaviors are consistent with the data in Refs.~{[}\onlinecite{Zn_later,paul}]
and Ref.~{[}\onlinecite{Muromachi12}], respectively. Temperatures
and $\delta_{0}$ are measured in units of $T_{c,0}$, which is the
SC transition temperature at perfect nesting and without SDW (for
pure SDW, the corresponding temperature is $T_{N,0}$). We used in
(a): $T_{N,0}/T_{c,0}=2$, $\delta_{2}/\left(2\pi T_{c,0}\right)=0.4$,
and impurity-scattering amplitudes $\Gamma_{0}=\Gamma_{\pi}=0.006\left(2\pi T_{c,0}\right)$;
and in (b): $T_{N,0}/T_{c,0}=4$, $\delta_{2}/\left(2\pi T_{c,0}\right)=0.8$,
and $\Gamma_{0}=\Gamma_{\pi}=0.012\left(2\pi T_{c,0}\right)$.}

\label{fig:1} 
\end{figure}

Our key results are summarized in Fig.~\ref{fig:1}, where we compare
the phase diagrams of FeSCs with and without impurities for two representative
values of the SDW and SC couplings and for on-site impurity potential
(i.e. equal intra-band and inter-band impurity scatterings). For one
set of parameters, Fig.~\ref{fig:1}(a), $T_{c}$ increases with
increasing disorder in the underdoped region. This behavior provides
an explanation for the experimental results of Refs.~{[}\onlinecite{Zn_later,paul}].
For the other set of parameters, Fig.~\ref{fig:1}(b), $T_{c}$ decreases
with disorder in the underdoped region, but with a smaller rate than
in the overdoped region. This behavior is consistent with the experiments
in Ref.~{[}\onlinecite{Muromachi12}].

\begin{figure}
\begin{centering}
\includegraphics[width=0.95\columnwidth]{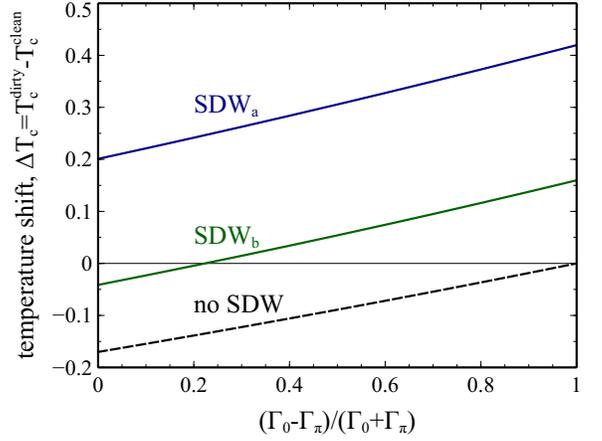} 
\par\end{centering}

\caption{(Color online) $\Delta T_{c}=T_{c}^{\mathrm{dirty}}-T_{c}^{\mathrm{clean}}$
(in units of $T_{c}^{\mathrm{clean}}$) as function of the ratio between
intra-band and inter-band impurity scattering amplitudes ($\Gamma_{0}$
and $\Gamma_{\pi}$, respectively). Line SDW$_{\mathrm{a}}$ is for
the parameters of Fig.~1(a) at a fixed $\delta_{0}/(2\pi T_{c,0})=0.2$
and line SDW$_{\mathrm{b}}$ is for the parameters of Fig.~1(b),
at a fixed $\delta_{0}/(2\pi T_{c,0})=0.6$. For both curves the system
is in the coexistence region and $(\Gamma_{0}+\Gamma_{\pi})/(2\pi T_{c}^{\mathrm{clean}})=0.01$.
The lower curve is for a pure superconductor, $M=0$. In the coexistence
region, $\Delta T_{c}$ is definitely positive when $\Gamma_{\pi}$
is small. For $\Gamma_{0}\sim\Gamma_{\pi}$, the behavior of $\Delta T_{c}$
is a result of the competition between the direct pair-breaking effect
of impurities, which tends to reduce $T_{c}$, and the suppression
of the SDW order parameter, which tends to increase $T_{c}$. }

\label{fig_2} 
\end{figure}

We also analyze the dependence of $T_{c}$ on the ratio of intra-band
and inter-band scattering rates (see Fig. \ref{fig_2}). For some
parameters, as those in Fig. \ref{fig:1}(a), $T_{c}$ increases for
any ratio of intra-band and inter-band impurity-scattering. For other
parameters, as those in Fig. \ref{fig:1}(b), $T_{c}$ increases when
intra-band scattering dominates and decreases when inter-band scattering
dominates. Even in the latter case, the rate at which $T_{c}$ decreases
with the strength of disorder is smaller than in a pure $s^{+-}$
superconductor.

\emph{The model}\textbf{\emph{~}}~~We follow earlier works~\cite{vvc,fs,eremin_chubukov2010}
and consider a minimal two-band model for the interplay between itinerant
SDW and $s^{+-}$ SC. In particular, we consider a circular hole pocket
at the center of the Fe-only Brillouin zone, and an elliptical electron
pocket displaced from the center by $\mathbf{Q}=\left(\pi,0\right)$
(or $\left(0,\pi\right)$). The non-interacting fermionic Hamiltonian
is \begin{equation}
\mathcal{H}_{0}=\sum_{\mathbf{k}\sigma}\varepsilon_{1,\mathbf{k}}c_{\mathbf{k}\sigma}^{\dagger}c_{\mathbf{k}\sigma}+\sum_{\mathbf{k}\sigma}\varepsilon_{2,\mathbf{k}}f_{\mathbf{k}\sigma}^{\dagger}f_{\mathbf{k}\sigma}\label{H0}\end{equation}
 where the operators $\hat{c}$ ($\hat{f}$) refer to electrons near
the hole (electron) pocket and the band dispersions are given by $\varepsilon_{1,\mathbf{k}}=-k^{2}/\left(2m\right)+\mu_{h}$
and $\varepsilon_{2,\mathbf{k}}=k_{x}^{2}/\left(2m_{x}\right)+k_{y}^{2}/\left(2m_{y}\right)-\mu_{e}$.
For small ellipticity $\left|m_{x}-m_{y}\right|\ll m$, the latter
can be conveniently parametrized by $\xi_{\mathbf{k}}=\left(\varepsilon_{1,\mathbf{k}}-\varepsilon_{2,\mathbf{k}}\right)/2\approx\mathbf{v}_{F}\cdot\left(\mathbf{k}-\mathbf{k}_{F}\right)$
and $\delta_{\mathbf{k}}=\left(\varepsilon_{1,\mathbf{k}}+\varepsilon_{2,\mathbf{k}}\right)/2=\left(\mu_{h}-\mu_{e}\right)/2+\left(k_{x}^{2}-k_{y}^{2}\right)\left(m_{x}-m_{y}\right)/4m\approx\delta_{0}+\delta_{2}\cos\varphi$.
We consider the interactions between the low-energy fermions in the
SDW (particle-hole) and SC (particle-particle) channels, as well as
their interaction with non-magnetic impurities. We first introduce
the SDW order parameter $M\propto\sum_{\mathbf{k},\sigma}\sigma\left\langle c_{\mathbf{k}\sigma}^{\dagger}f_{\mathbf{k}\sigma}\right\rangle $
and reduce the four-fermion SDW interaction to \begin{equation}
\mathcal{H}_{\mathrm{SDW}}=M\left(\sum_{\mathbf{k}}\sum_{\sigma=\pm1}\sigma c_{\mathbf{k}\sigma}^{\dagger}f_{\mathbf{k}\sigma}+\mathrm{h.c.}\right)\label{H_SDW}\end{equation}
 where $M$ is obtained self-consistently (see Eq. (\ref{self_cons_M})
below). Introducing the Nambu operators $\Psi_{\mathbf{k}}^{\dagger}=\left(c_{\mathbf{k}\uparrow}^{\dagger},c_{-\mathbf{k}\downarrow},f_{\mathbf{k}\uparrow}^{\dagger},f_{-\mathbf{k}\downarrow}\right)$,
the bare Green's function is expressed as \begin{equation}
\hat{\mathcal{G}}_{0}^{-1}=\left(\begin{array}{cc}
i\omega_{n}\hat{\tau}_{0}-\left(\xi+\delta\right)\hat{\tau}_{z} & M\hat{\tau}_{0}\\
M\hat{\tau}_{0} & i\omega_{n}\hat{\tau}_{0}+\left(\xi-\delta\right)\hat{\tau}_{z}\end{array}\right)\label{G0_bare}\end{equation}
 where $\hat{\tau}_{i}$ are Pauli matrices in Nambu space. In the
Born approximation, impurity scattering gives rise to the self-energy
correction $\hat{\Sigma}=n_{\mathrm{imp}}\sum_{\mathbf{k}}\hat{U}_{\mathbf{q-k}}\hat{\mathcal{G}}_{\mathbf{k}}\hat{U}_{\mathbf{k-q}}$,
where $n_{\mathrm{imp}}$ is the density of impurities, $\hat{\mathcal{G}}^{-1}=\hat{\mathcal{G}}_{0}^{-1}-\hat{\Sigma}$
is the renormalized Green's function, and $\hat{U}$ is the impurity
potential, which we decompose into an intra-band contribution $u_{0}$
and an inter-band contribution $u_{\pi}$: \begin{equation}
\hat{U}=\left(\begin{array}{cc}
u_{0}\hat{\tau}_{z} & u_{\pi}\hat{\tau}_{z}\\
u_{\pi}\hat{\tau}_{z} & u_{0}\hat{\tau}_{z}\end{array}\right)\label{U_imp}\end{equation}

\begin{figure}
\begin{centering}
\includegraphics[width=1\linewidth]{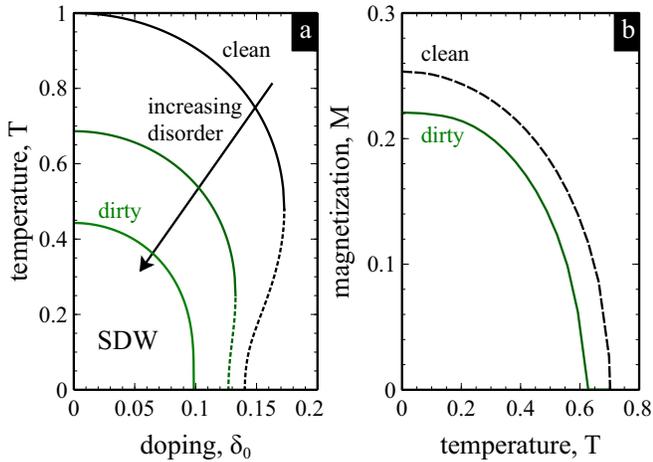} 
\par\end{centering}

\caption{(Color online) (a) SDW transition temperature (in units of $T_{N,0}$)
as function of doping $\delta_{0}$ (in units of $2\pi T_{N,0}$)
for the clean ($\gamma=0$) and dirty cases ($\gamma/2\pi T_{N,0}=0.06\,\mathrm{and}\,0.1$).
We set $\delta_{2}=0$, but the behavior at a finite $\delta_{2}$
is quite similar. The dashed lines denote metastable solutions of
the SDW gap equations, characteristic of the first-order character
of the transition. In the dirty case, the second-order SDW transition
extends to $T=0$ above a certain threshold of scattering amplitude.
(b) Magnetization $M$ (in units of $2\pi T_{N,0}$) as function of
temperature for $\delta_{0}/\left(2\pi T_{N,0}\right)=0.1$ and $\delta_{2}/\left(2\pi T_{N,0}\right)=0.2$
(same as in Fig.~\ref{fig:1}(a)). The dashed and solid lines are
for the clean and dirty cases, respectively (in the dirty case $\gamma/(2\pi T_{N,0})=0.006$).}

\label{fig_3} 
\end{figure}

To find $\hat{\mathcal{G}}$, we write it in the same form as Eq.
(\ref{G0_bare}), but with renormalized parameters $\tilde{\omega}_{n}$,
$\tilde{\delta}_{0}$, and complex $\tilde{M}$, with $\mathcal{G}_{13}^{-1}=\mathcal{G}_{31}^{-1}=\tilde{M}$
and $\mathcal{G}_{24}^{-1}=\mathcal{G}_{42}^{-1}=\tilde{M}^{*}$.
The parameter $\delta_{2}$ retains its bare value since $\left\langle \delta_{2}\cos2\varphi\right\rangle =0$
across the Fermi surface. Introducing the scattering amplitudes $\Gamma_{i}=\pi N_{F}n_{\mathrm{imp}}u_{i}^{2}$,
we obtain a set of self-consistent equations

\begin{eqnarray}
\tilde{\nu}_{n} & = & \nu_{n}+\gamma\left(\tilde{\nu}_{n}\tilde{\Pi}_{1}+i\tilde{\Pi}_{2}\right)\nonumber \\
\tilde{M} & = & M-\tilde{M}\gamma\tilde{\Pi}_{1}\nonumber \\
M & = & \lambda_{\mathrm{sdw}}T\sum_{n}\mathrm{Re}\left(\tilde{M}\tilde{\Pi}_{1}\right)\label{self_cons_M}\end{eqnarray}
 where $\nu_{n}=\omega_{n}+i\delta_{0}$,$\gamma=\Gamma_{0}+\Gamma_{\pi}$,
$M$ is the (real) SDW order parameter affected by impurities, and
$\lambda_{\mathrm{sdw}}$ is the SDW coupling constant. The Neel transition
temperature $T_{N,0}$ to a pure SDW phase at perfect nesting and
zero disorder is related to $\lambda_{\mathrm{sdw}}$ by $1/\lambda_{\mathrm{sdw}}=\ln(\Lambda/T_{N,0})$,
where $\Lambda$ is the high-energy cutoff. We emphasize that both
intra-band and inter-band impurity scattering affect $T_{N}$ and
the order parameter in the SDW channel. We also introduce \begin{equation}
\tilde{\Pi}_{1}\equiv\int\frac{d\varphi}{2\pi}\,\frac{1}{\tilde{\Omega}},\quad\tilde{\Pi}_{2}\equiv\int\frac{d\varphi}{2\pi}\,\frac{\cos2\varphi}{\tilde{\Omega}}\label{def_Pi1_2}\end{equation}
 with $\tilde{\Omega}=\sqrt{\tilde{M}^{2}+\left(\tilde{\nu}_{n}+i\delta_{2}\cos2\varphi\right)^{2}}$.
Note that $\tilde{\Pi}_{i}$ by itself depends on the renormalized
variables, i.e. Eqs.~(\ref{self_cons_M}) are non-linear self-consistent
equations.

The solution of the set (\ref{self_cons_M}) gives the Green's function
of a dirty SDW magnet, which then acts as a bare Green's function
for the SC system. Since we are only interested in $T_{c}$, we restrict
our analysis to the linearized SC gap equation in the presence of
impurities and a non-zero SDW order parameter: \begin{equation}
\frac{1}{T_{c}}=\lambda_{\mathrm{sc}}\sum_{n}\frac{\tilde{\Pi}_{3}}{1-\zeta\tilde{\Pi}_{3}}\label{self_cons_Delta}\end{equation}
 where $\zeta=\Gamma_{0}-\Gamma_{\pi}$, $1/\lambda_{\mathrm{sc}}=\ln(\Lambda/T_{c,0})$
is the coupling constant in the SC channel, $T_{c,0}$ is the superconducting
transition temperature without SDW and at perfect nesting, and \begin{equation}
\tilde{\Pi}_{3}=\int\frac{d\varphi}{2\pi}\left(\frac{1}{\tilde{\Omega}+\tilde{\Omega}^{*}}\right)\left(1+\frac{\left|\tilde{\nu}_{n}+i\delta_{2}\cos2\varphi\right|^{2}+\left|\tilde{M}\right|^{2}}{\left|\tilde{\Omega}\right|^{2}}\right)\label{Pi_3}\end{equation}
 Equation~(\ref{self_cons_Delta}) reduces to the gap equation of
an ordinary dirty $s^{+-}$ superconductor if we set $M=\tilde{M}=0$.
Alternatively, at perfect nesting, $\delta_{0}=\delta_{2}=0$, we
recover the results of Ref.~\cite{maxim}.

\textit{Results and comparison to experiments.}~~~ We first consider
the pure SDW state. In Fig. \ref{fig_3} we show that both the SDW
transition temperature $T_{N}$ and the order parameter $M$ are reduced
in the presence of impurities. This behavior is entirely expected,
since both intra-band and inter-band impurity scattering are detrimental
to SDW. A less obvious result is that impurities also affect the character
of the SDW transition at low $T$. In the clean case, the SDW transition
is first-order at low enough $T$.\cite{vvc} Impurities add additional
scattering and effectively shift $T\to T+\gamma$, extending the range
of the second-order transition to smaller temperatures $T$. Once
$\gamma$ exceeds a critical value $\gamma_{cr}$, the second-order
transition line extends down to $T=0$. The critical $\gamma_{cr}$
is obtained in a straightforward way by expanding the last equation
in (\ref{self_cons_M}) to order $M^{3}$ and verifying when the cubic
coefficient changes sign. For $\delta_{2}=0$, we obtain analytically
$\gamma_{cr}/\left(2\pi T_{N,0}\right)\approx0.08$, where $T_{N,0}$
is the SDW transition temperature at perfect nesting.

We now use the SDW results as input and solve Eq. (\ref{self_cons_Delta})
for $T_{c}$. To verify whether $T_{c}$ is reduced or enhanced with
increasing disorder, it is sufficient to consider small $\Gamma_{0}$
and $\Gamma_{\pi}$ and evaluate $\Delta T_{c}=T_{c}^{\mathrm{dirty}}-T_{c}^{\mathrm{clean}}$
to first order in $\Gamma_{i}/T_{N,0}$. The computations are tedious
but straightforward, so we skip the details and present our results.
The phase diagram in the presence of impurity scattering is shown
in Fig. \ref{fig:1} for on-site impurity potential ($\Gamma_{0}=\Gamma_{\pi}$)
and two ratios of $T_{N,0}/T_{c,0}$. We clearly see two different
types of behavior in the coexistence phase: $T_{c}$ either increases
when impurities are added, or decreases at a slow rate. This non-universal
behavior can be understood qualitatively: for a small $T_{N,0}/T_{c,0}$
(Fig.~\ref{fig:1}(a)) the effects of disorder on $T_{N}$ and $T_{c}$
are comparable, and the feedback on $T_{c}$ from the reduction of
$T_{N}$ overshadows the direct pair-breaking effect on $T_{c}$.
For a larger ratio $T_{N,0}/T_{c,0}$, the effect of disorder on SDW
gets relatively weaker, and the direct pair-breaking effect on $T_{c}$
prevails (more specifically, we estimate that $T_{c}$ increases with
disorder when the ratio $T_{N,0}/\left(\Gamma_{0}+\Gamma_{\pi}\right)$
is smaller than the ratio $T_{c,0}/\Gamma_{\pi}$). 

In Fig. \ref{fig_2} we plot $\Delta T_{c}=T_{c}^{\mathrm{dirty}}-T_{c}^{\mathrm{clean}}$
as a function of the ratio between intra-band and inter-band impurity
scattering amplitudes. We consider the two sets of parameters of Fig.
\ref{fig:1} with $\delta_{0}$ in the coexistence region and compare
them with the case when no SDW is present. In the latter, $\Delta T_{c}<0$
when $\Gamma_{\pi}$ is non-zero. We see that in the coexistence region
$\Delta T_{c}$ is definitely positive when $\Gamma_{0}/\Gamma_{\pi}$
is large enough, i.e. \textit{$T_{c}$ increases when impurities are
added into the coexistence state.} As expected, this increase is the
largest when $\Gamma_{\pi}$ vanishes, since in this limit impurities
are not pair-breaking, but still suppress $M$. When $\Gamma_{0}$
and $\Gamma_{\pi}$ are comparable, $T_{c}$ can either increase or
decrease, depending on parameters, but even when it decreases, the
rate of the decrease is smaller than that for a pure SC state.

Our results offer an explanation for the non-monotonic behavior of
$\Delta T_{c}=T_{c}^{\mathrm{dirty}}-T_{c}^{\mathrm{clean}}$ as function
of doping observed in Refs.~{[}\onlinecite{Zn_later,paul}] by adding
Zn to LaFeAs(O$_{1-x}$F$_{x}$) and Cu to Ba(Fe$_{1-x}$Co$_{x}$)$_{2}$As$_{2}$,
respectively. They also offer an explanation for the observation in
Ref.~{[}\onlinecite{Muromachi12}] that the addition of Zn to underdoped
Ba(Fe$_{1-x}$Co$_{x}$)$_{2}$As$_{2}$ leads to a decrease of $\Delta T_{c}$,
but at a slower rate than in the overdoped region. For this material,
the fact that Zn substitution leads to $\Delta T_{c}<0$ while Cu
substitution leads to $\Delta T_{c}>0$ may be due to different disorder
potentials associated with each dopant, leading to different ratios
$\Gamma_{0}/\Gamma_{\pi}$. It is also possible that the foreign element
not only acts as an impurity but also changes the electronic chemical
potential and/or the Fermi surface geometry.

We caution that the microscopic coexistence of SC and SDW orders has
been well-established for 122 compounds like Ba(Fe$_{1-x}$Co$_{x}$)$_{2}$As$_{2}$,
but this issue has not been settled for the 1111 systems like LaFeAs(O$_{1-x}$F$_{x}$).
Yet, even if SDW and SC phase-separate and occupy different parts
of the sample, we expect some of the physics described here to hold,
i.e. that impurity scattering on the one hand is pair-breaking and
on the other hand tends to increase $T_{c}$ by suppressing the competing
SDW phase.

\textit{Conclusions.}~~~ In summary, we showed that the different
behaviors of $T_{c}$ with impurity scattering observed in overdoped
and underdoped iron-based superconductors can be understood within
the $s^{+-}$ scenario for superconductivity. While in the overdoped
regime $T_{c}$ is quickly reduced with increasing impurity scattering,
in the underdoped regime there are two competing effects: the direct
pair-breaking by impurities, which reduces $T_{c}$, and the suppression
of the coexisting SDW order parameter, which increases $T_{c}$. We
demonstrated that, due to competition between these two effects, $T_{c}$
in the coexistence region either drops at a smaller rate or even increases
with increasing impurity concentration, in agreement with the experimental
data. We view this agreement as an evidence that the gap symmetry
in the iron pnictides is indeed $s^{+-}$.

We thank E. Bascones, S. Bud'ko, P. Canfield, F. Hardy, I. Eremin,
A. Kaminski, S. Maiti, Y. Matsuda, N. Ni, R. Prozorov, J. Schmalian,
M. Tanatar, A. Vorontsov, and Zhu-an Xu for useful discussions. R.M.F.
acknowledges the support from ICAM and NSF-DMR 0645461, as well as
the valuable support from the NSF Partnerships for International Research
and Education (PIRE) program OISE-0968226. M.G.V. and A.V.C. are supported
by NSF-DMR 0955500 and 0906953, respectively. R.M.F. and A.V.C. thank
the hospitality of the Aspen Center for Physics, where part of this
work has been done.


\begin{thebibliography}{10}
\bibitem{reviews} D. C. Johnston, Adv. Phys. \textbf{59}, 803 (2010);
J. Paglione and R. L. Greene, Nature Phys. \textbf{6}, 645 (2010).
P. J. Hirschfeld, M. M. Korshunov, and I. I. Mazin, Rep. Prog. Phys.
\textbf{74}, 124508 (2011); D. N. Basov and A. V. Chubukov, Nature
Phys. \textbf{7}, 241 (2011); P. C. Canfield and S. L. Bud'ko, Annu.
Rev. Cond. Mat. Phys. \textbf{1}, 27 (2010); H. H. Wen and S. Li,
Annu. Rev. Cond. Mat. Phys. \textbf{2}, 121 (2011); A. V. Chubukov,
Annu. Rev. Cond. Mat. Phys. \textbf{3}, 57 (2012); G. R. Stewart,
Rev. Mod. Phys. \textbf{83} 1589 (2011).

\bibitem{ARPES} H. Ding, P. Richard, K. Nakayama, K. Sugawara, T.
Arakane, Y. Sekiba, A. Takayama, S. Souma, T. Sato, T. Takahashi,
Z. Wang, X. Dai, Z. Fang, G. F. Chen, J. L. Luo and N. L. Wang, Euro.
Phys. Lett. \textbf{83} 47001 (2008); T. Kondo, A. F. Santander-Syro,
O. Copie, C. Liu, M. E. Tillman, E. D. Mun, J. Schmalian, S. L. Bud'ko,
M. A. Tanatar, P. C. Canfield, and A. Kaminski, Phys. Rev. Lett. \textbf{101},
147003 (2008); S. V. Borisenko, V. B. Zabolotnyy, D. V. Evtushinsky,
T. K. Kim, I. V. Morozov, A. N. Yaresko, A. A. Kordyuk, G. Behr, A.
Vasiliev, R. Follath, and B. Buchner, Phys. Rev. Lett. \textbf{105},
067002 (2010); T. Shimojima, F. Sakaguchi, K. Ishizaka, Y. Ishida,
T. Kiss, M. Okawa, T. Togashi, C.-T. Chen, S. Watanabe, M. Arita,
K. Shimada, H. Namatame, M. Taniguchi, K. Ohgushi, S. Kasahara, T.
Terashima, T. Shibauchi, Y. Matsuda, A. Chainani, and S. Shin, Science
\textbf{332}, 564 (2011).

\bibitem{maiti} S. Maiti, M. M. Korshunov, T. A. Maier, P. J. Hirschfeld,
and A. V. Chubukov, Phys. Rev. B \textbf{84}, 224505 (2011); Phys.
Rev. Lett. \textbf{107}, 147002 (2011).

\bibitem{magnetic} I. I. Mazin, D. J. Singh, M. D. Johannes, and
M. H. Du, Phys. Rev. Lett. \textbf{101}, 057003 (2008); K. Kuroki,
S. Onari, R. Arita, H. Usui, Y. Tanaka, H. Kontani, and H. Aoki, Phys.
Rev. Lett. \textbf{101}, 087004 (2008); J. Zhang, R. Sknepnek, R.
M. Fernandes, and J. Schmalian, Phys. Rev. B \textbf{79}, 220502(R)
(2009); A.F. Kemper, T.A. Maier, S. Graser, H-P. Cheng, P.J. Hirschfeld
and D.J. Scalapino, New J. Phys. \textbf{12}, 073030 (2010).

\bibitem{orbital} S. Onari and H. Kontani, Phys. Rev. Lett. \textbf{103},
177001 (2009); H. Kontani and S. Onari, Phys. Rev. Lett. \textbf{104},
157001 (2010); Y. Yanagi, Y. Yamakawa, and Y. Ono, Phys. Rev. B \textbf{81},
054518 (2010); T. Saito, S. Onari, and H. Kontani Phys. Rev. B \textbf{83},
140512(R) (2011).

\bibitem{dolgov2009} A. V. Chubukov, D. V. Efremov and I Eremin,
Phys. Rev. B \textbf{78}, 134512 (2008); O. V. Dolgov, A. A. Golubov,
D. Parker, New Journal of Physics, \textbf{11}, 075012 (2009); A.
B. Vorontsov, M. G. Vavilov, and A. V. Chubukov, Phys. Rev. B \textbf{79},
140507 (2009); Y. Bang, Europhys. Letters, \textbf{86}, 47001 (2009);
V.G. Kogan, Phys. Rev. B \textbf{80}, 214532 (2009); K. Nakamura,
R. Arita, and H. Ikeda, Phys. Rev. B \textbf{83}, 144512 (2011).

\bibitem{dhaka11} R. S. Dhaka, C. Liu, R. M. Fernandes, R. Jiang,
C. P. Strehlow, T. Kondo, A. Thaler, J. Schmalian, S. L. Bud'ko, P.
C. Canfield, and A. Kaminski, Phys. Rev. Lett. \textbf{107}, 267002
(2011); K. Kirshenbaum, S.R. Saha, S. Ziemak, T. Drye, and J. Paglione,
arXiv:1203.5114

\bibitem{Zn_early} Y. K. Li, X. Lin, Q. Tao, C. Wang, T. Zhou, L.
Li, Q. Wang, M. He, G. Cao, and Z. A. Xu, New J. Phys. \textbf{11},
053008 (2009).

\bibitem{Zn_later} Y. K. Li, J. Tong, Q. Tao, C. Feng, G. Cao, W.
Chen, F. C. Zhang, and Z. A. Xu, New J. Phys. \textbf{12}, 083008
(2010).

\bibitem{Muromachi10} Y. F. Guo, Y. G. Shi, S. Yu, A. A. Belik, Y.
Matsushita, M. Tanaka, Y. Katsuya, K. Kobayashi, I. Nowik, I. Felner,
V. P. S. Awana, K. Yamaura, and E. Takayama-Muromachi, Phys. Rev.
B \textbf{82}, 054506 (2010).

\bibitem{paul} N. Ni, A. Thaler, J. Q. Yan, A. Kracher, E. Colombier,
S. L. Bud'ko, and P. C. Canfield, Phys. Rev. B \textbf{82}, 024519
(2010).

\bibitem{goldman_Cu} M. G. Kim \emph{et al}, arXiv:1204.1538

\bibitem{Muromachi12} J. Li, Y. Guo, S. Zhang, Y. Tsujimoto, X. Wang,
C. I. Sathish, S. Yu, K. Yamaura, and E. Takayama-Muromachi, Solid
State Commun. \textbf{152}, 671 (2012).

\bibitem{vvc} A. B. Vorontsov, M. G. Vavilov, and A. V. Chubukov,
Phys. Rev. B \textbf{81}, 174538 (2010), Phys. Rev. B \textbf{84},
140502 (2011).

\bibitem{fs} R. M. Fernandes, D. K. Pratt, W. Tian, J. Zarestky,
A. Kreyssig, S. Nandi, M. G. Kim, A. Thaler, N. Ni, P. C. Canfield,
R. J. McQueeney, J. Schmalian, and A. I. Goldman, Phys. Rev. B \textbf{81},
140501 (2010); R. M. Fernandes and J. Schmalian, Phys. Rev. B \textbf{82},
014520 (2010); \emph{ibid} Phys. Rev. B \textbf{82}, 014521 (2010).

\bibitem{kulikov1984} M. T. Rice, Phys. Rev. B \textbf{2}, 3619 (1970);
N. Kulikov and V.V.Tugushev, Sov. Phys. Usp. \textbf{27}, 954 (1984),
{[}Usp.Fiz.Nauk \textbf{144} 643 (1984)]; V. Cvetkovic and Z. Tesanovic,
Europhys. Lett. \textbf{85}, 37002 (2009).

\bibitem{comm} Two of us recently used the fact that both SDW and
SC are affected by impurities to argue that the phase diagram in case
when doping acts as a non-magnetic impuritry is the same as when doping
modifies the band structure.\cite{maxim}

\bibitem{maxim} M.G. Vavilov and A. V. Chubukov, Phys. Rev. B \textbf{84},
214521 (2011).

\bibitem{eremin_chubukov2010} I. Eremin and A.V. Chubukov, Phys.
Rev. B \textbf{81}, 024511 (2010). 
\end{thebibliography}
\end{document}